\documentstyle[epsfig]{elsart}
\begin{document}
\begin{frontmatter}

\title{Breakup time scale studied in the 
8 GeV/c $\pi^-$ + $^{197}$Au reaction}

\author[HIL]{L.~Pienkowski\thanksref{1}}, 
\author[IUCF]{K.~Kwiatkowski\thanksref{2}}, 
\author[IUCF]{T.~Lefort}, 
\author[IUCF]{W.-c.~Hsi\thanksref{3}}, 
\author[IUCF]{L.~Beaulieu\thanksref{4}},  
\author[RUS]{A.~Botvina}, 
\author[ARG]{B.~Back}, 
\author[MAR]{H.~Breuer},  
\author[BRO]{S.~Gushue}, 
\author[BUR]{R.G.~Korteling},  
\author[TEX]{R.~Laforest\thanksref{5}}, 
\author[TEX]{E.~Martin}, 
\author[TEX]{E.~Ramakrishnan}, 
\author[BRO]{L.P.~Remsberg}, 
\author[TEX]{D.~Rowland}, 
\author[TEX]{A.~Ruangma}, 
\author[IUCF]{V.E.~Viola}, 
\author[TEX]{E.~Winchester}, 
\author[TEX]{S.J.~Yennello}

\address[HIL]{Heavy Ion Laboratory, Warsaw University, ul. Pasteura 5a, 
              02-093 Warszawa, Poland} 
\address[IUCF]{Department of Chemistry and IUCF, Indiana University,
               Bloomington, IN 47405, USA}
\address[RUS]{Institute for Nuclear Research, Russian Academy of Science, 
              117312 Moscow, Russia}
\address[ARG]{Physics Division, Argonne National Laboratory, 
              Argonne IL 60439, USA}
\address[MAR]{Department of Physics, University of Maryland, College Park, 
              MD 20740, USA}
\address[BRO]{Chemistry Department, Brookhaven National Laboratory, 
              Upton, NY 11973, USA}
\address[BUR]{Department of Chemistry, Simon Fraser University, Burnaby,
              BC V5A IS6, Canada}
\address[TEX]{Department of Chemistry \& Cyclotron Laboratory, Texas A\&M
              University, College Station, TX 77843, USA}

\thanks [1] {E-mail: pienkows@slcj.uw.edu.pl}
\thanks [2] {Present address: Los Alamos National Laboratory, 
    Los Alamos, NM 87545}
\thanks [3] {Present address: 7745 Lake Street, 
    Morton Grove IL 60053}
\thanks [4] {Present address: Departement de Radio-oncologie, 
    Universitaire de Laval, Quebec, Canada}
\thanks [5] {Present address: Barnes Hospital, Washington 
    University, St. Louis, MO  63130}

\begin{abstract}

 Experimental data from the reaction of an 8.0 GeV/c 
$\rm{\pi^-}$ beam incident on a $\rm{^{197}}$Au target have been 
analyzed in order to investigate the integrated breakup time scale
for hot residues. Alpha-particle energy spectra and particle angular 
distributions supported by 
a momentum tensor analysis suggest that at large excitation energy, 
above 3-5 MeV/nucleon, light-charged particles are emitted prior to 
or at the same 
time as the emission of the heavy fragments. Comparison with the 
SMM and GEMINI models is presented. A binary fission-like 
mechanism fits the experimental data at low excitation energies, 
but seems unable to reproduce the data at excitation energies 
above 3-5 MeV/nucleon. 

\end{abstract}

\end{frontmatter}

\newpage

 During the past decade a major effort has been directed toward studies 
of the decay of hot nuclei. Experiments with energetic light 
projectiles offer a unique perspective into decay processes 
at excitation energies up to about 9 MeV/nucleon.
The formation of highly-excited heavy 
residues in such reactions is less complicated than in heavy-ion 
reactions, since only a single hot source is formed and small angular 
momenta and little compression are involved \cite{Cug87,Bot88}.
The subsequent evolution of the emission process for heavy fragments 
as a function of excitation energy is of major concern in understanding 
the decay mechanisms of hot nuclear matter.

 At low excitation energies fission fragments are emitted near the end 
of the sequential decay chain. At an excitation energy of 
about 3 MeV/nucleon the emission is delayed by about $10^3$ fm/c for 
the asymmetric split and this delay is much larger for symmetric division 
\cite{Hil89}. Recently it has been shown that even at 
$E^*/A$ of about 4 MeV/nucleon, fission fragments from the decay of hot 
gold-like nuclei are observed with the probability of about 
20-40\% \cite{Jan99}. The onset of different decay processes involving 
the prompt emission of multiple fragments is predicted to occur at 
excitation energies above 4-5 MeV/nucleon \cite{Bon95,Gro97}.  

The experimental evidence for the transition from a sequential binary 
decay picture to a simultaneous breakup mechanism is discussed in this 
letter. Some other signals of the expected transition have recently 
been presented \cite{Bea00,Lef00} on the basis of the experimental 
data set analyzed for the purpose of this letter. 
The Statistical Multifragmentation Model (SMM) \cite{Bon95,Bot90} 
is one of the ``explosive'' models and the predictions from this model 
are compared with the experimental data. A freeze-out volume of 
$V=3\cdot V_0$ (radius $R=1.44\cdot R_0$), without any extra expansion 
energy \cite{Lef00}, which describes the $Z\ge 6$ multiplicity and 
charge distributions at large excitation energies, was assumed 
for all SMM calculations. Results from the sequential decay model 
GEMINI \cite{GEMINI}, including decay time scales and modified 
to calculate the trajectories in a mutual Coulomb field \cite{GAW97} 
are also presented. As input for all simulations, the mass and charge 
of the hot residual nucleus at a given excitation energy were based 
on event reconstruction procedure of the experimental data \cite{Bea99}.


 Experiment E900a was performed at the AGS Accelerator at Brookhaven 
National Laboratory. A secondary 8 GeV/c beam of tagged $\pi^-$ 
irradiated a 2~mg/cm$^2$ gold target. The ISiS $4\pi$ detector array 
detects charged particles and fragments with charge up to $Z\approx 16$ 
units and $E/A\approx 1-92$ MeV \cite{Kwiat95}. The experimental methods 
and previous results can be found in \cite{Bea99,Lef99}. In this letter 
attention is concentrated on events that detect at least one heavy 
Intermediate Mass Fragment, IMF, specifically fragments with 
$Z_{IMF}=8-16$ charge units.

\begin{figure}
\centerline{\epsfig{figure=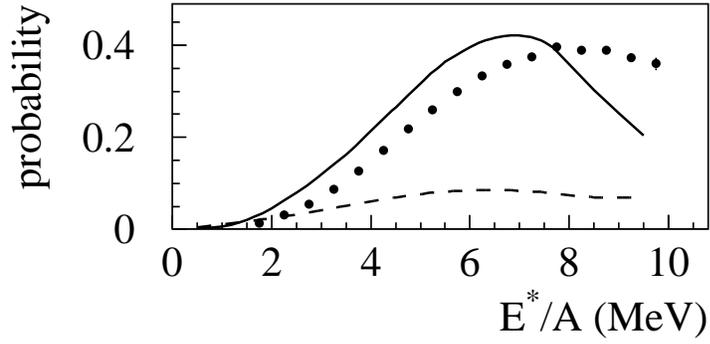,width=10.0cm}}
\caption{
Dots present the probability to detect an event with a heavy IMF, 
$Z_{IMF}=8-16$ and solid (dashed) line presents the SMM (GEMINI) 
model predictions filtered with the experimental detection efficiency. 
An initial angular momentum of hot nucleus $L=20\hbar$ was assumed 
for GEMINI model calculations.}
\label{Fig1}
\end{figure}

The probability for events with heavy IMFs is presented as a function 
of $E^*/A$ in Fig.~\ref{Fig1}. It is observed that the probability 
is very low at low $E^*/A$, increases noticeably in the excitation energy 
range 3-6 MeV/nucleon and reaches about 40\% at about $E^*/A\approx 7-8$ 
MeV/nucleon. Taking into account 74\% detection efficiency, it can be  
concluded that the selected events are typical events at the excitation 
energy of about 6-8 MeV/nucleon, where the detected average number 
of thermal charged particles is about 12-15, the number of fast cascade 
particle is 10-15, based on a sample of $10^5$ events in this $E^*/A$ range. 
The thermal particles and fragments were selected by imposing a cutoff 
energy that accepted only fragments below 30 MeV for $Z=1$ and 
$9\cdot Z+40$ MeV for heavier fragments \cite{Kwiat98}. SMM and GEMINI 
models predict for selected events that up to $E^*/A$=7-9 MeV/nucleon 
the heavy fragment multiplicity with charge $Z \ge 8$ is two or slightly 
larger than two. The second, mainly heavier, fragment than that detected 
is too slow to be efficiently detected by ISiS. The event reconstruction 
procedure gives an estimation of the total missing charge that is 
consistent with a single missing heavy fragment which is equivalent 
to an IMF for $E^*/A > 7$ MeV/nucleon. It is observed on Fig.~\ref{Fig1} 
that experimental data are in good agreement with the SMM model predictions 
up to $E^*/A \approx 8$ MeV/nucleon. The GEMINI model with a density 
parameter a=A/10 MeV$^{-1}$ underestimates the probability to detect 
heavy IMFs above excitation energies 3-5 MeV/nucleon. 


 In the 8.0 GeV/c $\rm{\pi^- +^{197}Au}$ reaction the center-of-mass 
system, CM, is very close to the laboratory system, LAB, 
$v_{source}\approx 0.01 c$. It is observed that the thermal particles 
and fragments are emitted almost isotropically \cite{Lefbor99,Gol96,Hsi99}, 
consistent with emission from a thermal-like system. The data were 
also checked to insure that fast particles were not correlated with 
the heavy fragments and their spectra were independent of the angle 
relative to the heavy fragment axis. As shown in model 
calculations~\cite{Cug87,Bot88}, the initial spin of the hot nuclei 
is low, so for the studies of the thermal decay of the hot nuclei 
the beam axis doesn't define any special direction. Therefore, 
the direction of the heaviest detected IMF in the LAB system was 
used as the reference axis for further studies.

\begin{figure}
\centerline{\epsfig{file=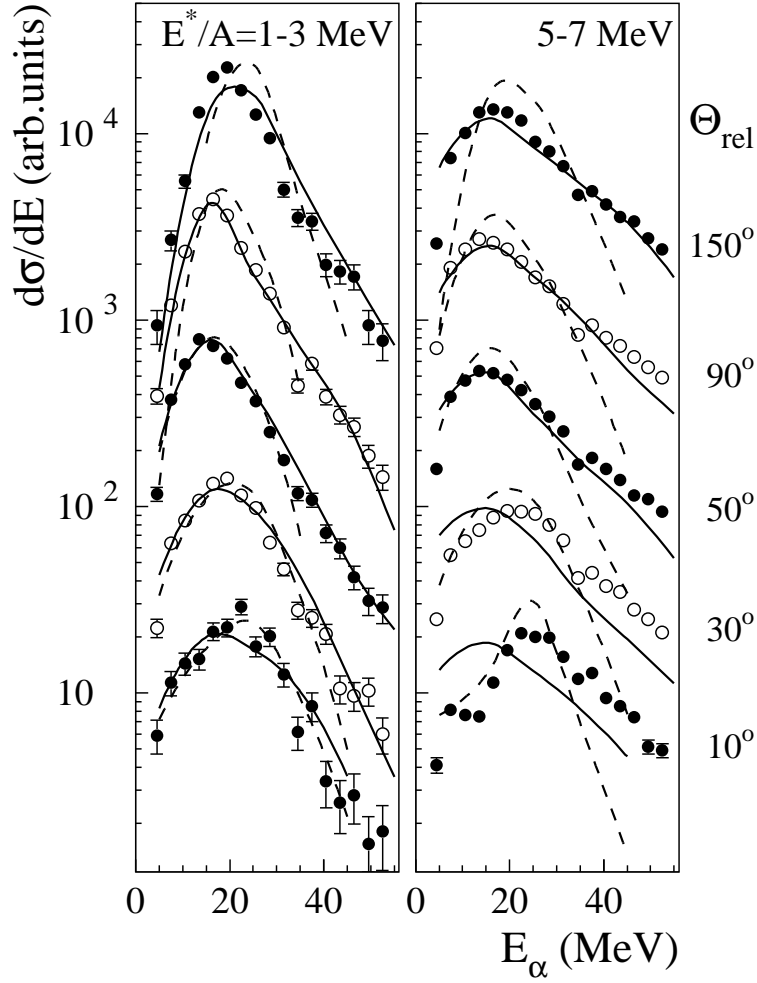,width=10.0cm}}
\vspace{0.2cm}
\caption{Alpha-particle energy spectra for selected excitation energy 
and angular bins relative to the heaviest fragment, $Z_{IMF}=8-16$, 
emission axis, $\Theta_{rel}$. Solid (dashed) lines present the SMM 
(GEMINI) prediction filtered with the experimental detection efficiency.
The simulation results are normalized to the experimental data assuming 
the same cross section at a given angle.}
\label{Fig2}
\end{figure}

 Alpha-particle kinetic energy spectra are presented in Fig.~\ref{Fig2},  
for a few selected angles, $\Theta_{rel}$, relative to the axis of the 
heaviest detected fragment $Z_{IMF}=8-16$. On the left side of Fig.~\ref{Fig2} 
the spectra at $E^*/A$=1-3 MeV/nucleon are presented. Over most of the 
angular range the shape of the alpha-particle energy spectra doesn't 
change much. However, the thermal alpha-particle average kinetic 
energy, $<E_{kin}>$, has a maximum value at small- and large-angles, 
$<E_{kin}>\approx 23$ MeV. The minimum is observed at 50 deg, 
$<E_{kin}>\approx 20$ MeV.
 The GEMINI model and SMM model (running at $E^*/A\le 3$ MeV/nucleon 
essentially as a sequential evaporation model) fit the shape 
of the spectra, however, predict a more pronounced evolution than 
the experimental data.

 The spectra at $E^*/A$=5-7 MeV/nucleon are presented on the right side 
of the Fig.~\ref{Fig2}. It is important to note that the multifragmentation 
model SMM reproduces the slope of the spectra up to about 60 MeV. The 
GEMINI model predicts much steeper spectra. It is also observed that 
at angles close to the axis defined by a heavy fragment, the alpha-particle 
kinetic energy spectra have a different shape than the spectra at the larger 
angles. The average kinetic energy is $<E_{kin}>\approx 29$ MeV, 
compared to a nearly constant value of $<E_{kin}>\approx 24$ MeV 
at the angles between 50 and 170 deg. Similar evolution of the energy 
spectra is also observed for protons and lithium isotopes, i.e. at large 
excitation and small angles, $\Theta_{rel}$, the kinetic energy spectra 
are shifted towards higher energy. The modified GEMINI model \cite{GAW97} 
predicts a shift similar to the observed one. Tracing the calculations, 
it was found that at $E^*/A$=5-7 MeV/nucleon GEMINI predicts a short, 
exponentially distributed time for heavy fragment separation, 
on average about 100 fm/c. According to the model, most of the 
alpha-particles are emitted from the hot fragments. The modified 
version of GEMINI model \cite{GAW97} shows that preferentially 
low energy alpha-particles are suppressed towards small angles, 
$\Theta_{rel}$. Particles emitted from the complementary fragment 
towards the selected fragment are deflected by the Coulomb field 
to larger angles $\Theta_{rel}$. This result of the model indicates 
that secondary decay of the hot fragments (emitted promptly after 
thermalization) is able to explain the kinetic energy shift at small 
angles, in despite of all doubts concerning the validity of GEMINI 
model at large excitation energies. 

 The SMM model doesn't show the observed energy shift of alpha-particles 
at small angles. This failure seems to be related to the difficulties 
to define a set of parameters at the freeze-out point configuration. 
In this context it should be noticed that the version of the SMM model 
used assumes that secondary decay occurs after the heavy fragments have 
been fully accelerated, at infinite separation distance, taking into 
account only two-body Coulomb interaction. 


 It is interesting to discuss the possibility to describe the shift 
of alpha-particle kinetic energy at large excitation energy, and 
at small angles, $\Theta_{rel}$, by a fission-like decay scenario. 
This requires that the hot nucleus survives as a single self-bound object, 
the heavy fragments separate after a certain delay time, and the 
alpha-particles are mainly emitted prior to fragment separation, 
possibly from a strongly deformed system. One can envision two 
touching hot spherical nuclei at rest, which would resemble a largely 
deformed system. In a such a case the alpha-particle energy depends only 
slightly on the emission angle. Particle emission from the neck at the 
scission configuration seems to be the only mechanism to enhance 
the yield of low energy particles emitted perpendicular to the fragments.
However, it is observed that the average emitted thermal alpha-particle 
multiplicity at $E^*/A$=5-7 MeV/nucleon is about 5 and to suppress low 
energy alpha-particles at small angles, the emission from the neck region 
would have to be a dominant decay process. In other words, the hot nucleus 
should emit multiple particles mainly from the strongly deformed system, 
primarily from the neck region, and not from the tip region. Hence, 
it seems to be difficult to describe such behavior by a fission-like 
decay mechanism. 

\begin{figure}
\centerline{\epsfig{file=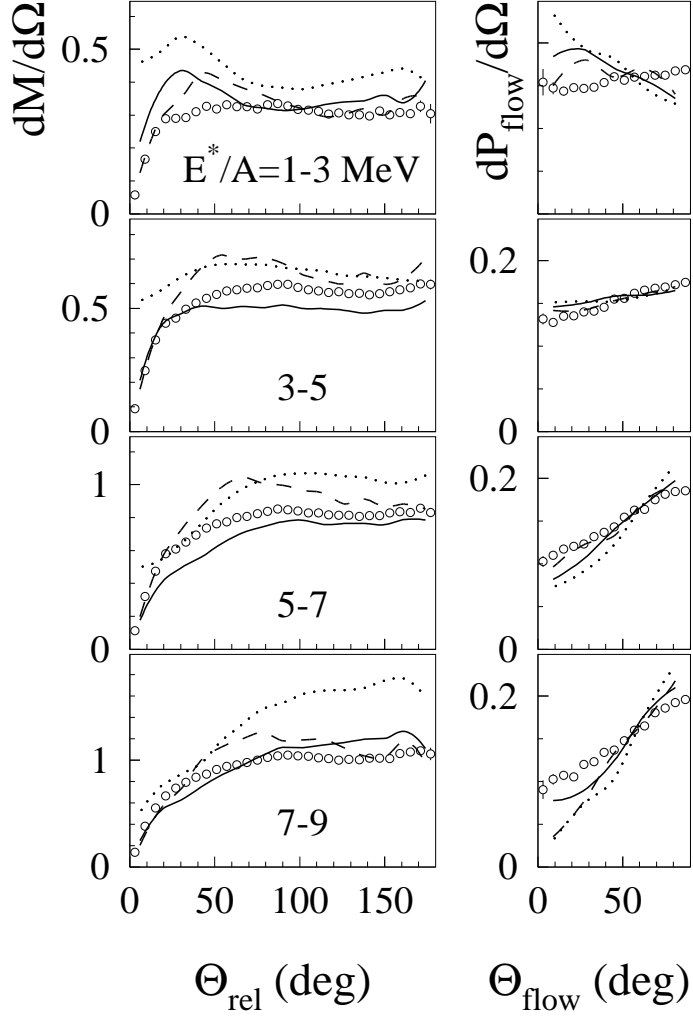,width=9.0cm}}
\vspace{0.5cm}
\caption{Laboratory angular distributions with respect to the axis defined 
by the emission of the heaviest detected IMF, $Z_{IMF}=8-16$ charge units, 
for the selected excitation energy bins. Left side: Average multiplicity 
of the detected thermal charged particles with $Z=1-7$ charge units as 
a function of $\Theta_{rel}$. Right side: probability distribution of 
the momentum flow angle, $\Theta_{flow}$. Solid dots present the experimental 
data. Solid (dotted) lines present the SMM model predictions filtered 
(non filtered) with the experimental detection efficiency. Dashed lines 
are from GEMINI model predictions filtered with the experimental detection 
efficiency.}
\label{Fig3}
\end{figure}

The angular multiplicity distributions of thermal particles and light 
fragments ($Z=1-7$) relative to the emission axis of a $Z_{IMF}=8-16$ 
fragment are shown on the left side of Fig.~\ref{Fig3}. 
The particle multiplicity distributions are plotted for selected 
excitation energy bins, along with SMM model and GEMINI model predictions. 
In all such angular distribution plots the decrease at angles below 
15 deg reflects the ISiS angular granularity.
The GEMINI model and SMM model at $E^*/A$ below 3 MeV/nucleon predict 
a bump at small angles due to particles emitted from the accelerated 
fragments. The absence of enhanced emission along the heavy IMF axis 
in the experimental data indicates that the accelerated fragments 
have lower excitation energies than predicted by the models. 
Such a picture is consistent with the binary fission scenario, i.e. 
the fragments are emitted at the end of the sequential 
decay chain, in a good agreement with refs. \cite{Hil89,Jan99}. 

 The angular distributions at larger $E^*/A$ show that the probability 
of detecting a particle or fragment close to the selected fragment is 
suppressed, and suggests that most of the charged particles are emitted 
almost simultaneous with the selected fragment emission. In other words, 
the Coulomb field of the heavy fragment makes a cone-shadow not accessible 
to the other particles emitted at the same time. Consistent with this 
picture, the SMM model predictions filtered with the detection efficiency 
procedure describe the experimental data well. At the very large 
excitation energy, $E^*/A$=7-9 MeV/nucleon, the GEMINI model also 
fits the data due to the very short time scale predicted by the model: 
the heavy fragment relative separation time of about 75 fm/c, 
and the first neutron emission decay time of about 6 fm/c. 
However, the GEMINI predictions should be taken with the 
special care at large excitation energies. The predicted decay 
time at $E^*/A$=7-9 MeV/nucleon is shorter than the time to pass 
the distance equal to the nucleus diameter with a speed of light. 
This indicates that the decay time predicted by the GEMINI model 
is shorter than the energy relaxation time, which contradicts 
to the model assumption. 

 In order to investigate systematically the total breakup time of the 
system and the hypothesis of simultaneous decay, a momentum tensor 
analysis \cite{Cug83} was performed on the charged particles. 
The tensor was built for each event from all detected thermal charged 
particles and light fragments with $Z=1-7$ charge units. Each element 
of the tensor, $Q_{ij}$, was calculated as:

\begin{equation} \label{equ:tensor}
Q_{ij}=\sum_{\nu}^{N}{\frac{p_i^{(\nu)}\cdot p_j^{(\nu)}}{p^{(\nu)}}}
\end{equation}

where, $N$ is the total number of particles in an event, $p^{(\nu)}$ is the 
particle momentum in the center-of-mass system of $N$ particles 
with $Z=1-7$ and $p_{i,j}^{(\nu)}$ are the particle momentum Cartesian 
coordinates. The tensor was diagonalized and the angle between the vector 
associated with the largest eigenvalue and the emission direction of 
the selected heaviest fragment, $Z_{IMF}=8-16$, was defined as momentum 
flow angle, ${\Theta}_{flow}$. This method permits study of the 
correlations between the heavy fragments and lighter products 
of the reaction. 

 The distributions of the angle ${\Theta}_{flow}$ are presented on the 
right side of the Fig.~\ref{Fig3}. The experimental data are almost 
isotropically distributed at low excitation energy but develop a maximum 
at 90 deg that becomes increasingly pronounced with increasing excitation 
energy. An isotropic ${\Theta}_{flow}$ distribution assumption describes 
more than 95\% of events at $E^*/A$=1-3 MeV/nucleon but less than 60\% 
of events at $E^*/A$=7-9 MeV/nucleon. These experimental results should 
be discussed in a frame of different decay scenarios.

 If particles are emitted isotropically, before fragment separation, 
then ${\Theta}_{flow}$ has also an isotropic distribution. According to 
the binary fission scenario, this should be the case for the experimental 
data at $E^*/A$ below 3 MeV/nucleon. The small enhancement of events, 
below 5\%, with  ${\Theta}_{flow}$ greater than zero seems to be due 
to particle emission perpendicular to the fission axis, perhaps from 
the neck region at the scission configuration. 

 If there are additional particles emitted from the accelerated fragments, 
then the momentum ellipsoid should be elongated in the direction of 
the heavy fragment emission and the ${\Theta}_{flow}$ distribution 
should be focused at small angles (the kinematical focusing effect). 
This behavior is shown for low excitation energy by the GEMINI and 
SMM predictions, corresponding to the angular distributions on the 
left side of the Fig.~\ref{Fig3}. 

 On the other hand, if the associated charged particles are emitted at the 
same time as the heavy fragment, then preferential Coulomb focusing in a 
direction perpendicular to the selected heavy fragment emission axis is 
expected. In such a case the momentum ellipsoid is elongated perpendicular 
to the heavy fragment emission direction and ${\Theta}_{flow}$ should 
present a maximum at 90 deg. This focusing effect normal to the heavy 
fragment axis gives evidence that at high excitation energy the Coulomb 
focusing effect is stronger than any kinematical focusing effects. The 
prompt decay scenario, including the Coulomb focusing effect, is contained 
in the SMM model. Thus, at large excitation energy the SMM model describes 
the ${\Theta}_{flow}$ distribution well. The modified version of the 
GEMINI model, with very short emission times after thermalization 
also fits the the ${\Theta}_{flow}$ data at large excitation energy.

 The integrated time scale for breakup of the thermal-like system 
at excitation energies above 5-7 MeV/nucleon can be estimated from 
the experimental data. At high excitation energy the observed stronger 
Coulomb focusing relative to kinematical focusing (right side of 
the Fig.~\ref{Fig3}) means that fewer particles are emitted after 
the acceleration time compared to the number of particles emitted 
within the first 200 fm/c after fragment formation. The modified 
GEMINI model gives an estimation that more than about 75\% of 
particles emitted after fragment separation would have to be emitted 
within the first 200 fm/c to create stronger Coulomb focusing effect 
than kinematical focusing effect. The light fragment, $Z\le 7$, 
angular distributions (Fig.~\ref{Fig3}) and alpha-particle kinetic 
energy spectra (Fig.~\ref{Fig2}) indicate that not many particles 
can be emitted after the thermalization but before heavy fragment 
separation. Taking into account these experimental results, it can be 
concluded that at excitation energies above 5-7 MeV/nucleon the 
disintegration of the system typically takes place within a total 
time of approximatively 200 fm/c after thermalization. 
This short lifetime is in a good agreement with the results obtained 
from the IMF-IMF intensity-interferometry analysis based on the same 
experimental data \cite{Bea00}. 

 The presented results are obtained for events with detected 
heavy IMF, $Z=8-16$, and IMF-IMF interferometry analysis was 
based on the set of events with detected two light IMFs, 
$Z_{1,2}=4-9$. At a large excitation energies, above 5-7 MeV/nucleon 
these two set of events represent a large, almost complementary 
set of events, and cover together more than 70\% of detected events. 
However, it seems that the method based on the sample of events with 
the splits closer to a symmetric split is a better choice looking 
for the experimental evidences of the expected transformation from 
the shape deformation decay mode to the prompt decay mode. Moreover, 
the set of events with detected heavy fragments covers a large range 
of excitation energy, the fission-like events were already successfully 
used in the past \cite{Hil89,Hil92} to study the decay time scale at low 
excitation energy, and the IMF-IMF interferometry technique has access 
only to a marginally low fraction of events at low excitation energy 
due to a very low IMF emission probability.


In conclusion, it is shown that for hot nuclei at excitation energies 
below 3-5 MeV/nucleon the experimental data are consistent with a 
sequential binary breakup scenario \cite{Hil89,Jan99}, i.e. the heavy 
fragments are emitted at the end of a sequential decay chain.
The fission-like decay scenario above $E^*/A$ of about 6 MeV/nucleon seems 
unable to fit the experimental data, because it would require emission 
of too many particles from the neck of the scission configuration.
 At excitation energies above 3-5 MeV/nucleon, charged-particle emission 
appears to occur nearly simultaneously with the emission of heavy fragments, 
$Z_{IMF}=8-16$. The integrated breakup time of the system estimated for the 
experimental data is approximatively 200 fm/c. This observation is 
in agreement with the delay times from IMF-IMF small angle correlation 
studies \cite{Bea00} and Statistical Multifragmentation Model (SMM) 
predictions. 

\textbf{Acknowledgements:} This work was supported by the U.S. 
Department of Energy and National Science Foundation, the National and 
Engineering Research Council of Canada, Grant No. P03B 048 15 of the 
Polish State Committee for Scientific Research, Indiana University Office 
of Research and the University Graduate School, Simon Fraser University 
and the Robert A. Welch Foundation.

\newpage

\newpage

\newpage

\end{document}